\documentclass{PoS}

\usepackage{psfrag}
\usepackage{subfigure}

\newlength{\figwidth}
\newlength{\halffigwidth}
\title{Unquenching the Landau Gauge Lattice Propagators and the Gribov Problem}

\ShortTitle{Unquenching the Landau Gauge Lattice Propagators and the Gribov Problem}

\author{\speaker{Paulo J. Silva} and Orlando Oliveira\\
       Centro de F\'isica Computacional, Universidade de Coimbra, Portugal\\
        E-mail: \email{psilva@teor.fis.uc.pt}, \email{orlando@teor.fis.uc.pt} }

\abstract{The gluon and ghost propagators are computed using both quenched and
dynamical configurations for the same lattice spacings. The Wilson
fermions are simulated at several quark masses. Furthermore, the effect
of the Gribov copies is evaluated for all sets of configurations.}

\FullConference{The XXVIII International Symposium on Lattice Field Theory\\
                 June 14-19,2010\\
                 Villasimius, Sardinia Italy}

\begin{document}

\section{Introduction and motivation}

It is well-known that the Faddeev-Popov (FP) procedure \cite{FaPo67} can not 
be applied to the non-perturbative regime of Quantum Chromodynamics (QCD), 
due to the so-called Gribov copies pro\-blem. The FP trick assumes a unique 
solution of the gauge fixing condition on each set of gauge related 
configurations. Gribov \cite{Gribov} has shown, for the Landau and Coulomb 
gauges, that there are more than one configuration satisfying the gauge 
condition on each gauge orbit. Furthermore, it is not possible 
to define a local continuous gauge fixing condition free of 
Gribov copies \cite{singer78, killingback83}.

Many (quenched) lattice simulations have confirmed the existence 
of Gribov copies \cite{latgribov}, and the influence of this issue 
on Landau gauge gluon and ghost propagators have been studied 
(see, for example, \cite{gribovcopies, gribov07}). The simulations show
that for the gluon propagator the influence of Gribov copies is 
null or small, whereas for the ghost propagator the effect is larger. 

In this paper we recover the computation of gluon and ghost propagators on 
dynamical configurations, including an investigation of the Gribov copies.
Similar studies have been performed by other groups while ago \cite{dyngluon}
but, to the best of our knowledge, no investigation on the influence of the 
Gribov copies in dynamical simulations is found in the literature.

\section{Lattice setup}

%Gluon propagator: dynamical vs. quenched
\begin{figure}[p]
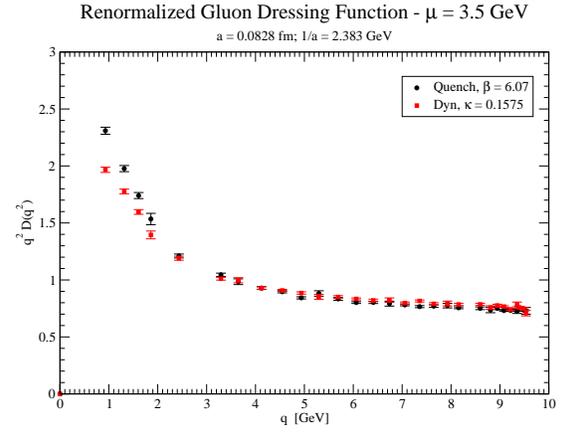
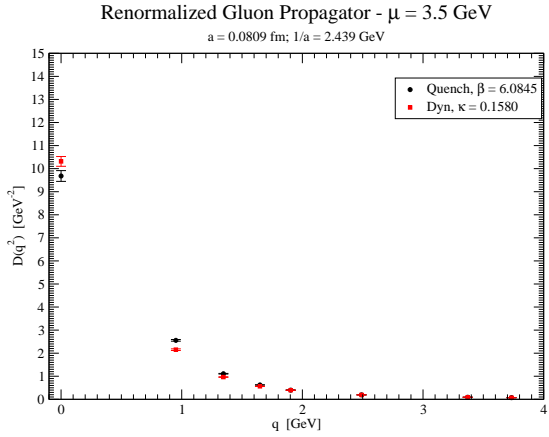
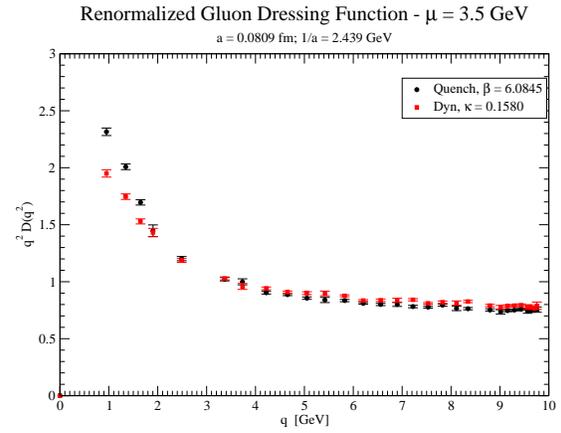
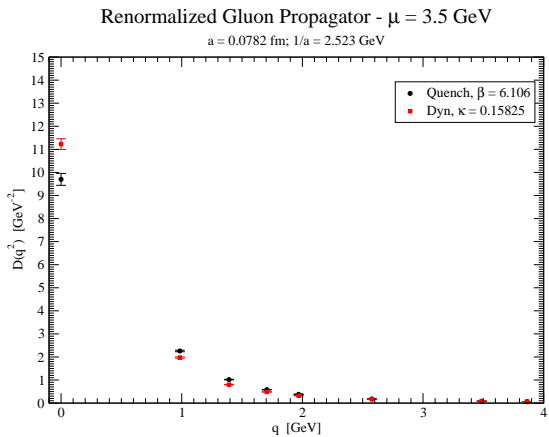
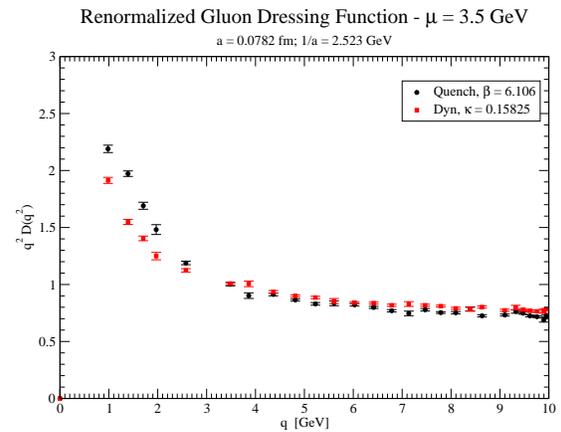
    
  \subfigure[Gluon propagator at $a=0.0828$fm.]{
  \begin{minipage}[b]{0.45\textwidth}
    \centering
    \includegraphics[origin=c,angle=0,width=\halffigwidth]{Dmu3.5GeV_beta6.07.eps}
  \end{minipage} } \hfill
  \subfigure[Gluon dressing function at $a=0.0828$fm. ]{
  \begin{minipage}[b]{0.45\textwidth}
    \centering
    \includegraphics[origin=c,angle=0,width=\halffigwidth]{q2Dmu3.5GeV_beta6.07.eps}
  \end{minipage} }\vspace*{0.8cm}
  \subfigure[Gluon propagator at $a=0.0809$fm. ]{
  \begin{minipage}[b]{0.45\textwidth}
    \centering
    \includegraphics[origin=c,angle=0,width=\halffigwidth]{Dmu3.5GeV_beta6.0845.eps}
  \end{minipage} } \hfill
  \subfigure[Gluon dressing function at $a=0.0809$fm.]{
  \begin{minipage}[b]{0.45\textwidth}
    \centering
    \includegraphics[origin=c,angle=0,width=\halffigwidth]{q2Dmu3.5GeV_beta6.0845.eps}
  \end{minipage} }\vspace*{0.8cm}
  \subfigure[Gluon propagator at $a=0.0782$fm. ]{
  \begin{minipage}[b]{0.45\textwidth}
    \centering
    \includegraphics[origin=c,angle=0,width=\halffigwidth]{Dmu3.5GeV_beta6.106.eps}
  \end{minipage} } \hfill
  \subfigure[Gluon dressing function at $a=0.0782$fm.]{
  \begin{minipage}[b]{0.45\textwidth}
    \centering
    \includegraphics[origin=c,angle=0,width=\halffigwidth]{q2Dmu3.5GeV_beta6.106.eps}
  \end{minipage} }\vspace*{0.8cm}
  \caption{Dynamical gluon propagators and dressing functions \textit{versus} quenched counterparts. }
\label{dyngluon}
\end{figure}

\begin{table}[b]
	\begin{center}
  	\begin{tabular}{ccccccc}
    	\hline
\multicolumn{3}{c}{Quenched}  & \multicolumn{4}{c}{Dynamical Wilson}\\
$\beta$ & a(fm) & $\#$ conf. & $\kappa$ & $\pi$ mass (MeV) & $q$ mass (MeV) & $\#$ conf. \\
\hline
6.07   & 0.0828 &   60      & 0.1575   &   665        & 66       &  71 \\
6.0845 & 0.0809 &   66      & 0.1580   &   485        & 34       &  71 \\
6.106  & 0.0782 &   62      & 0.15825  &   385        & 22       &  71 \\
\hline
\end{tabular}
\caption{Lattice setup. \label{latsetup}}
\end{center}	
\end{table}

In this paper we will consider $16^4$ configurations. 
Three different sets of dynamical configurations for Wilson gauge action 
at $\beta=5.6$, using Wilson fermions, with 
different $\kappa$'s, see table \ref{latsetup}, were considered. 
For the dynamical ensembles, the lattice spacing have been computed 
using the Sommer scale $r_0=0.5$ fm. Data for $ar_0$, $m_q$ and $m_{\pi}$ 
have been taken from \cite{urbach}. The $\beta$ values for the corresponding 
quenched ensembles have been computed using \cite{necco} so that they have 
the same lattice spacing as the dynamical counterparts. Quenched simulations 
have been performed using MILC code \cite{milc}, whereas dynamical 
simulations have been performed using Chroma software library \cite{chroma}; 
the MD integrator used in the HMC simulations has been tuned using the 
techniques described in \cite{poissonery}.

Details on the lattice Landau gauge fixing and computation of the 
propagators can be found in \cite{gribov07}. The lattice data for the 
gluon propagator has been renormalized by requiring 
\begin{equation}
D_{R}(q^2)|_{q^2={\mu}^2}= \frac{1}{\mu^2}=Z_{R}D_{Lat}(\mu^2)
\end{equation}
at $\mu=3.5$ GeV. In what concerns the ghost propagator, due to the very small 
statistical errors, we were unable to fit the high momenta in order to perform 
the renormalization. Given that we observed no difference in the UV between 
quenched and unquenched propagators, we just show the bare data.

In order to evaluate the effect of Gribov copies, we have generated 100 
copies per gauge configuration. We will consider, besides a random copy, 
the copies which have the largest and smallest value of the gauge fixing 
functional.

\section{Results}

\begin{figure}[t]
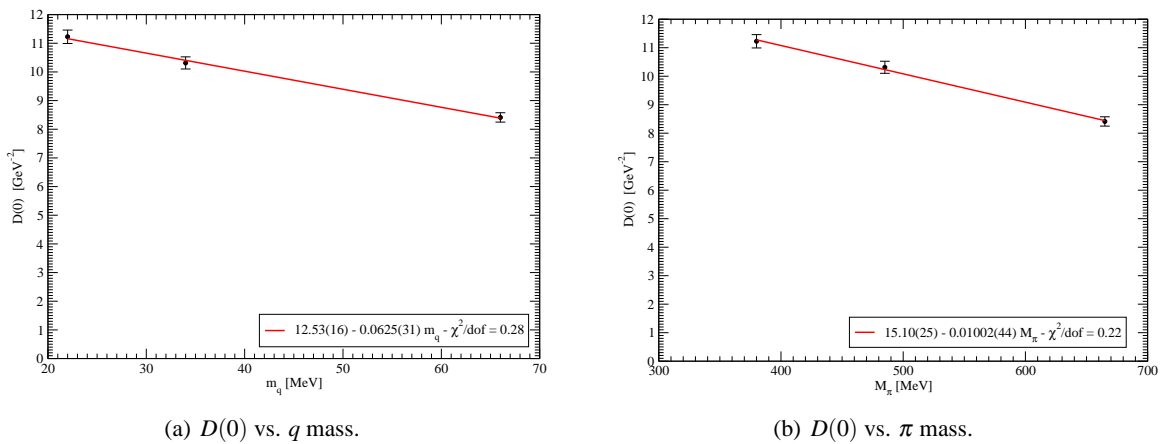
    
  \subfigure[$D(0)$ vs. $q$ mass.]{
  \begin{minipage}[b]{0.45\textwidth}
    \centering
    \includegraphics[origin=c,angle=0,width=\halffigwidth]{D0mu3.5GeV_Mq.eps}
  \end{minipage} } \hfill
  \subfigure[$D(0)$ vs. $\pi$ mass.]{
  \begin{minipage}[b]{0.45\textwidth}
    \centering
    \includegraphics[origin=c,angle=0,width=\halffigwidth]{D0mu3.5GeV_Mpi.eps}
  \end{minipage} }\vspace*{0.8cm}
  \caption{Zero momentum gluon propagator as a function of quark and pion masses. }
\label{D0dyn}
\end{figure}

In figure \ref{dyngluon} the gluon propagator computed 
with dynamical fermions is compared with the quenched propagator. 
Figure \ref{dyngluon} shows clearly that the effect of dynamical quarks 
is mainly seen in the infrared region. With the exception of the zero 
momentum, the propagator for the dynamical configurations is suppressed
in the infrared. On the other hand, $D(0)$ from the dynamical ensembles 
is suppressed for heavier quarks, and enhanced for lighter quarks. 
Furthermore, $D_{dyn}(0)$ follows a linear behaviour 
as a function of the quark and pion masses -- see figure \ref{D0dyn}.

\begin{figure}[b]
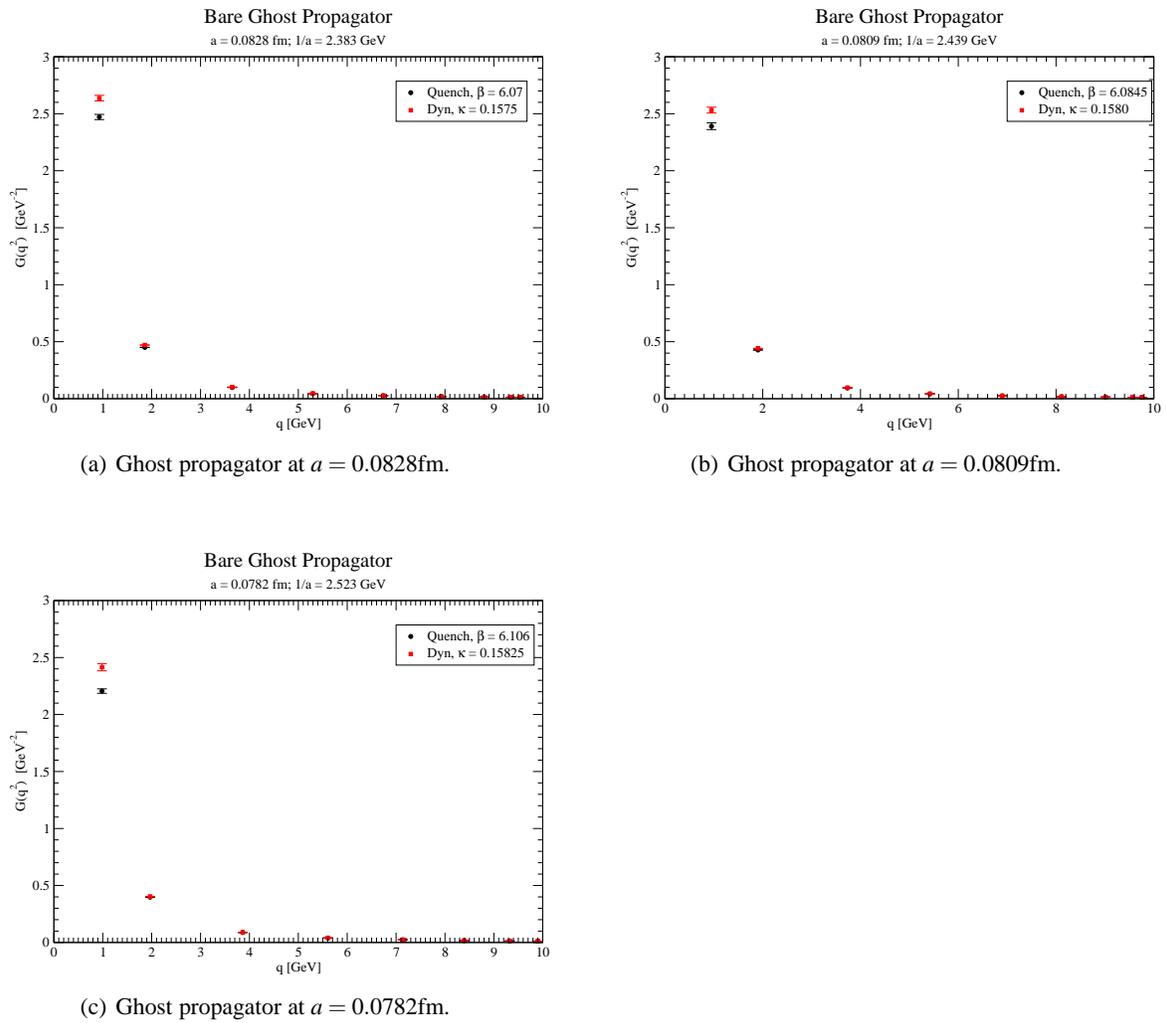
    
  \subfigure[Ghost propagator at $a=0.0828$fm.]{
  \begin{minipage}[b]{0.45\textwidth}
    \centering
    \includegraphics[origin=c,angle=0,width=\halffigwidth]{ghost_bare_b6.07.eps}
  \end{minipage} } \hfill
  \subfigure[Ghost propagator at $a=0.0809$fm.]{
  \begin{minipage}[b]{0.45\textwidth}
    \centering
    \includegraphics[origin=c,angle=0,width=\halffigwidth]{ghost_bare_b6.0845.eps}
  \end{minipage} }\vspace*{0.8cm}
  \subfigure[Ghost propagator at $a=0.0782$fm.]{
  \begin{minipage}[b]{0.45\textwidth}
    \centering
    \includegraphics[origin=c,angle=0,width=\halffigwidth]{ghost_bare_b6.106.eps}
  \end{minipage} } \vspace*{0.8cm}
\caption{Dynamical bare ghost propagators \textit{versus} quenched counterparts. }
\label{ghost}
\end{figure}

In what concerns the ghost propagator, see figure \ref{ghost}, the effect 
of dynamical configurations is again more important at low momenta, 
with the dynamical quarks enhancing the propagator 
in the infrared region, if compared with quenched results. 
The data also shows that the enhancement goes with the quark mass, 
i.e. the larger the quark mass, the larger is the enhancement.

%Ghost propagator: quenched vs. dynamical

\begin{figure}[t]
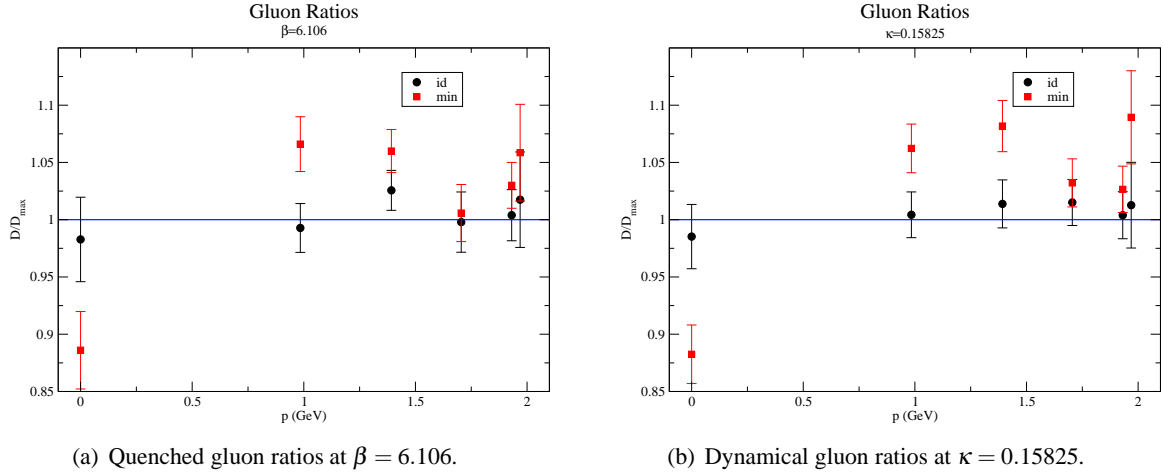
  
  \subfigure[Quenched gluon ratios at $\beta=6.106$.]{
  \begin{minipage}[b]{0.45\textwidth}
    \centering
    \includegraphics[origin=c,angle=0,width=\halffigwidth]{ratio_b6.106.eps}
  \end{minipage} } \hfill
  \subfigure[Dynamical gluon ratios at $\kappa=0.15825$.]{
  \begin{minipage}[b]{0.45\textwidth}
    \centering
    \includegraphics[origin=c,angle=0,width=\halffigwidth]{ratio_k15825.eps}
  \end{minipage} }\vspace*{0.8cm}
  \caption{Gluon ratios.}
\label{gribovgluon}
\end{figure}

\begin{figure}[b]
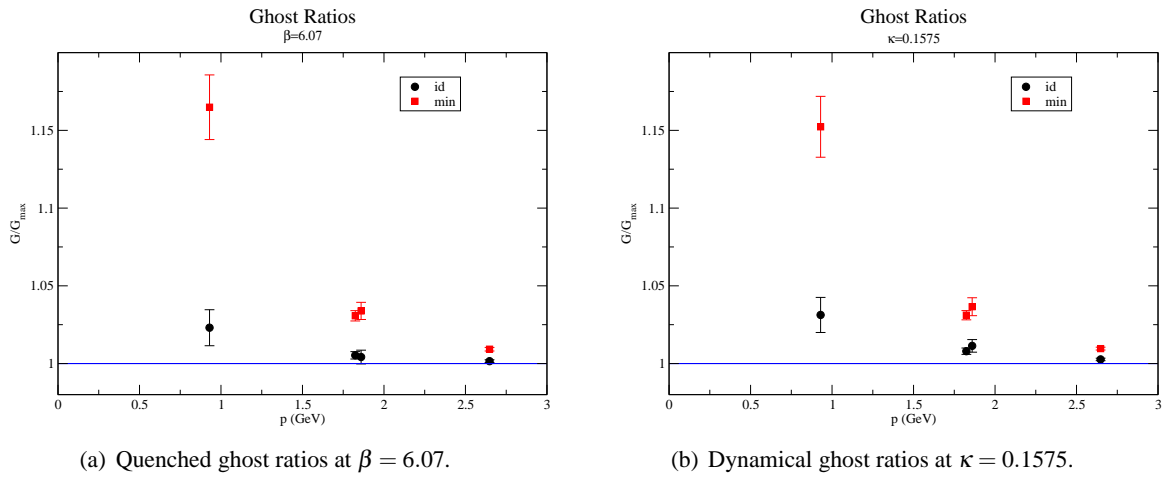
    
  \subfigure[Quenched ghost ratios at $\beta=6.07$.]{
  \begin{minipage}[b]{0.45\textwidth}
    \centering
    \includegraphics[origin=c,angle=0,width=\halffigwidth]{ghostratio_b6.07.eps}
  \end{minipage} } \hfill
  \subfigure[Dynamical ghost ratios at $\kappa=0.1575$. ]{
  \begin{minipage}[b]{0.45\textwidth}
    \centering
    \includegraphics[origin=c,angle=0,width=\halffigwidth]{ghostratio_k1575.eps}
  \end{minipage} }\vspace*{0.8cm}
  \caption{Ghost ratios. }
\label{gribovghost}
\end{figure}

The influence of the Gribov copies in the propagators is resumed in figures 
\ref{gribovgluon} and \ref{gribovghost}. Indeed, the pattern observed for 
the quenched and unquenched theories is similar: no or small effects 
for the gluon propagator; larger effects for the ghost propagator. 
Note that the Gribov effects in the ghost propagator are seen over a large 
range of momenta, with the propagator being suppressed for larger F maxima.

%Gluon Ratios
%Ghost ratios

\section{Conclusions}

We have presented a first study of the effect of Gribov copies in Landau gauge gluon and ghost propagators computed from dynamical configurations. The results show that the effect is very similar for quenched and dynamical propagators. Furthermore, we have revisited the effects of dynamical quarks in gluon and ghost propagators. 

\acknowledgments

P. J. Silva acknowledges support from FCT via grant SFRH/BPD/40998/2007. O. Oliveira acknowledges support from FAPESP. The authors acknowlegde financial support from FCT under contracts PTDC/FIS/100968/2008 and CERN/FP/109327/2009.

\end{document}